%% file: eprint.tex
\documentclass[12pt]{article}
\usepackage{graphicx}

\usepackage{xspace} 

\usepackage{amsmath} 
\usepackage{amssymb}
\usepackage{amsfonts}
\usepackage{upgreek} 

\usepackage{hyperref}    
\usepackage[all]{hypcap} 
\usepackage{cite} 
\usepackage{mciteplus}

\input{my-symbols}
\input{lhcb-symbols-def}

\usepackage{ifthen} 
\newboolean{articletitles}
\setboolean{articletitles}{false} 

\newcommand\pubnumber{}
\newcommand\pubdate{\today}

\def\heidelberg{Ruprecht-Karls-Universitaet Heidelberg, Physikalisches Institut\\
Im Neuenheimer Feld 226, 69120 Heidelberg, GERMANY}

\def\Title#1{\begin{center} {\Large #1 } \end{center}}
\def\Author#1{\begin{center}{ \sc #1} \end{center}}
\def\Address#1{\begin{center}{ \it #1} \end{center}}

\newcommand\pubblock{\rightline{\begin{tabular}{l} \pubnumber\\
         \pubdate  \end{tabular}}}
\newenvironment{Abstract}{\begin{quotation}  }{\end{quotation}}
\newenvironment{Presented}{\begin{quotation} \begin{center} 
             PRESENTED AT\end{center}\bigskip 
      \begin{center}\begin{large}}{\end{large}\end{center} \end{quotation}}
\def\Acknowledgements{\bigskip  \bigskip \begin{center} \begin{large}
             \bf ACKNOWLEDGEMENTS \end{large}\end{center}}

\input econfmacros.tex

\begin{document}
\begin{titlepage}
\pubblock

\vfill
\Title{Measurement of \CP observables\\ in semileptonic decays at \lhcb}
\vfill
\Author{ Lucia Grillo \\ on behalf of the \lhcb collaboration}
\Address{\heidelberg}
\vfill
\begin{Abstract}
\lhcb has recorded large samples of semileptonic \B decays. These provide the possibility to study \CP violation effects in the \Bs and \Bd systems. Decay-time-integrated or decay-time-dependent asymmetries between charge conjugated final states probe \CP violation in $B^{0}_{(s)}$ mixing through the measurement of the parameter  \asl. These measurements rely on data-driven techniques to control possible detection asymmetries.
\end{Abstract}
\vfill
\begin{Presented}
8th International Workshop on the CKM Unitarity Traingle (CKM 2014)\\
8-12 September 2014, Vienna, Austria
\end{Presented}
\vfill
\end{titlepage}
\def\thefootnote{\fnsymbol{footnote}}
\setcounter{footnote}{0}

\section{Introduction}

Mixing of neutral $B$ mesons is defined by the flavour-changing $B \rightarrow \Bbar$ and $\Bbar \rightarrow B$ transitions, which in the Standard Model are described by higher order loop diagrams. 
Measurements of the so called mixing asymmetry in  \B-meson systems probe whether the oscillation of the meson into its antimeson is different from its revers process. In a flavour-specific $B$ meson decay, the final state $f$ unambiguously identifies the flavour of the $B$ meson at the decay.
Due to the high statistics of the samples, it is experimentally convenient to use semileptonic decays. 
\begin{equation}
\label{eq:asl}
 \asl =  \frac{\Gamma(\overline{B} \rightarrow B \rightarrow f) - \Gamma(B \rightarrow \overline{B} \rightarrow \overline{f})}{\Gamma(\overline{B} \rightarrow B \rightarrow f) + \Gamma(B \rightarrow \overline{B} \rightarrow \overline{f})}
\end{equation}
The flavour-specific semileptonic asymmetry is defined in terms of the partial decay rates. Semileptonic asymmetries have a precise and very small Standard Model prediction: $\asls = (1.9 \pm 0.3) \times 10^{-5}$ in the case of the \Bs system and $\asld = (-4.1\pm 0.6) \times 10^{-4}$ for the \Bd system \cite{Lenz:2006hd}\cite{Lenz:2011ti} ,  but they could be enhanced by new physics contributions. 
Measurements of the single asymmetries \asls and \asld, performed by the B-factories and the \dzero experiment, show agreement with the SM predictions \cite{Jaffe:2001hz} \cite{Nakano:2005jb} \cite{Aubert:2004xga} \cite{BaBar_asld_new} \cite{Lees:2013sua} \cite{Abazov:2012zz} \cite{Abazov:2012hha} \cite{BaBar_asld_new}. On the other hand, a deviation of 3.6 $\sigma$ has been reported by the \dzero experiment in the inclusive charge asymmetry in events with same charge dimuons \cite{Abazov:2013uma}.  More precise measurements of these quantities are needed to clarify the picture.
At hadron colliders the identification of the flavor of the $B$ mesons at production is rather inefficient. For this reason at LHCb, in order to access \asls and \asld, the decay-time-integrated or the decay-time-dependent asymmetry between the \CP conjugated final states, regardless the production flavor of the \B meson, is measured.
The measured final state asymmetry as a function of the \B decay time, relates to \asl as follows:
\begin{equation}
\label{eq:Ameas}
A_{\rm meas}(t) = \frac{\Gamma(f,t) - \Gamma(\bar{f},t)}{\Gamma(f,t) + \Gamma(\bar{f},t)} = \frac{a_{\rm sl}^q}{2}+A_{\rm D} - \Big(A_{\rm P}+ \frac{a_{\rm sl}^q}{2}\Big) \frac{{\rm cos(}\Delta m_q t{\rm )}}{{\rm cosh(} \Delta \Gamma_q t/2{\rm)}}
\end{equation}
with $q = s, d$. $\Delta m_q$ is the mixing frequency of the $B^{0}_{(s)}$ meson, $\Delta \Gamma_q$ the decay width difference between its heavy and light mass eigenstates, and $\Gamma_q$ the average decay width.  $\AP \equiv [\sigma(\Bzsb) - \sigma(B^{0}_{(s)})]/[\sigma(\Bzsb) + \sigma(B^{0}_{(s)})]$ represents the production asymmetry of the $B^{0}_{(s)}$ and \Bzsb mesons at \lhcb, in the selected kinematic phase space. \AD $\equiv [\epsilon(f) - \epsilon(\bar{f})]/[\epsilon(f) + \epsilon(\bar{f})]$ is the asymmetry between the $CP$ conjugated final states due to detector and reconstruction efficiencies. \\
In this talk, I report on measurements of \asls and \asld performed at LHCb, using datasets corresponding to an integrated luminosity of $1~ {\rm fb^{-1}}$ and $3~ {\rm fb^{-1}}$. The first measurement of \asld, focuses on the crucial experimental challenges: the ability to disentangle the \CP violating asymmetry from the \Bd production asymmetry and the final state particles detection asymmetries.

\section{Measurement of  \CP violation in $B^{0}_s -\Bsb$ mixing}
In this analysis, the time-integrated asymmetry between $D_s^{+} [ \to \phi \pi^{+}] X \mu^{-}\nu$ and $D_s^{-} [ \to \phi \pi^{-}] X \mu^{+}\nu$ decays, with $X$ representing possible associated particles, is measured. This asymmetry, calculated with the time-integrated decay rates, relates to the $CP$ violating asymmetry in $B^{0}_{s}$ mixing, \asls,  similarly to Eq. \ref{eq:Ameas}.
After integrating over decay time, the time dependent term in Eq. \ref{eq:Ameas} is suppressed to a neglible level by the large value of \dms. The production asymmetry is expected to be at the percent level ~\cite{LHCb-PAPER-2014-042}.
The measured final state asymmetry then relates to \asls as $A_{\rm meas} = \asls \cdot 0.5 + A_{\rm D}$.\\
The raw asymmetry is calculated from the signal yields, determined from the $K^{+}K^{-}\pi^{+}$ and $K^{+}K^{-}\pi^{-}$  invariant mass distributions. It is corrected by the detection asymmetry and the asymmetry due to background decays not well separated from the signal decays in the data sample.
Backgrounds include prompt charm production, fake muons associated with real $D^{+}_{s}$ mesons produced in b-hadron decays, and $B \to D D_s$ decays where the $D$ hadron decays semileptonically. 
To measure the relative $\pi^{+}$ and $\pi^{-}$ detection efficiencies, the ratio of partially reconstructed and fully reconstructed $D^{*+} \to \pi^{+} D^{0}$ with $D^{0} \to K^{-}\pi^{+}\pi^{+}(\pi^{-})$ decays is used, where the pion in brackets is not required in the partial reconstruction. From the tracking asymmetry of the pion, and the difference in kinematic phase space between the muon and the pion and between the two kaons, the asymmetry due to different tracking efficiencies is accessed.\\
The muon trigger and identification asymmetry are evaluated using a $J/\psi \to \mu^{+}\mu^{-}$ sample, where both muons are reconstructed in the tracking system, but with at least one muon without trigger and identification requirements.\\ 
The result obtained for \asls with a data sample corresponding to an integrated luminosity of $1~ \rm fb^{-1}$ is \cite{LHCb-PAPER-2013-033}:
\begin{equation}
\label{eq:asls_result}
 a_{\rm sl}^{s} =  (-0.06 \pm 0.50({\rm stat}) \pm 0.36({\rm syst}))\%\\
 \end{equation}

\section{Measurement of  \CP violation in $B^{0} - \Bzb$ mixing}
\begin{figure}[htb]
\centering
\includegraphics[height=2.2in]{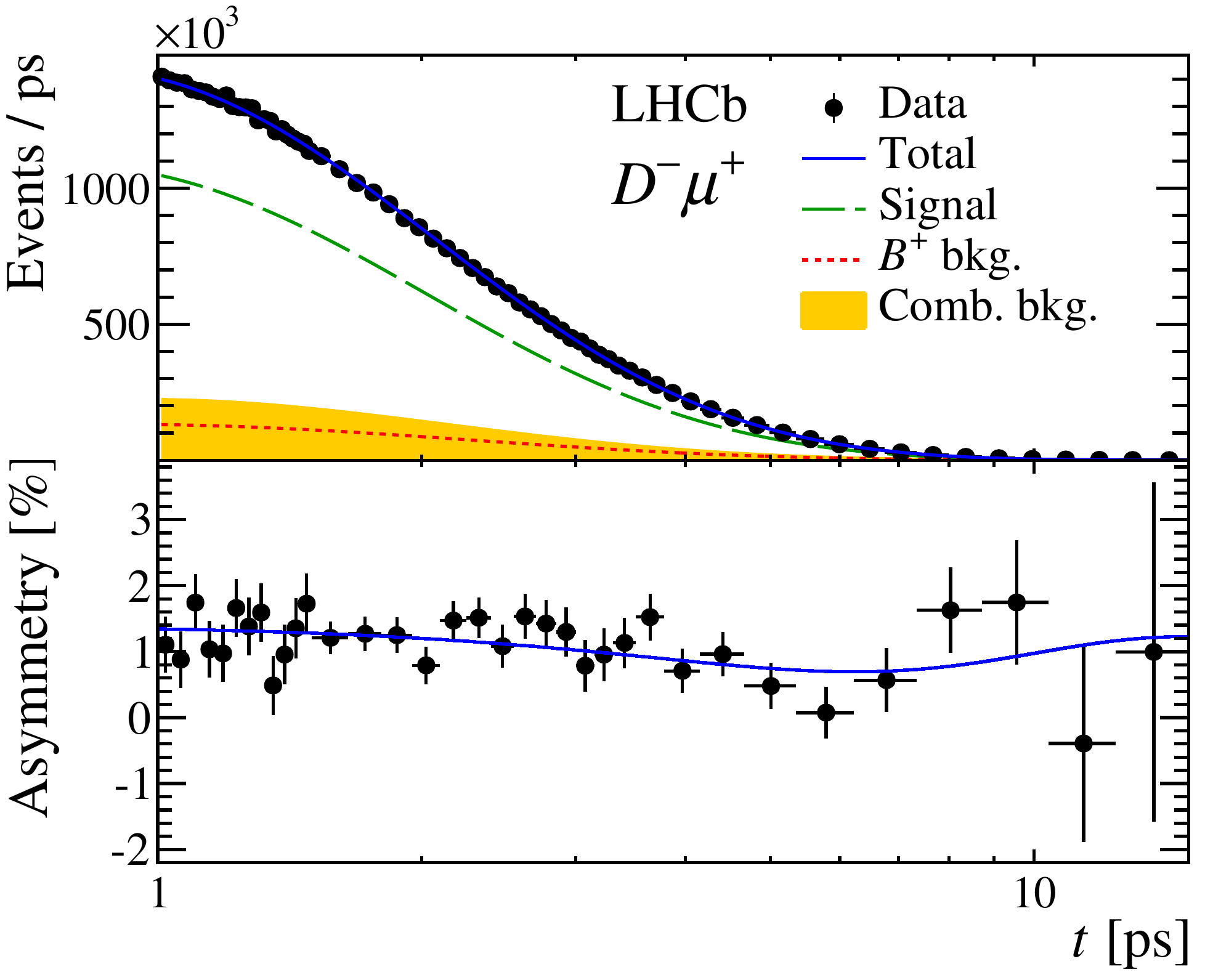}
\includegraphics[height=2.2in]{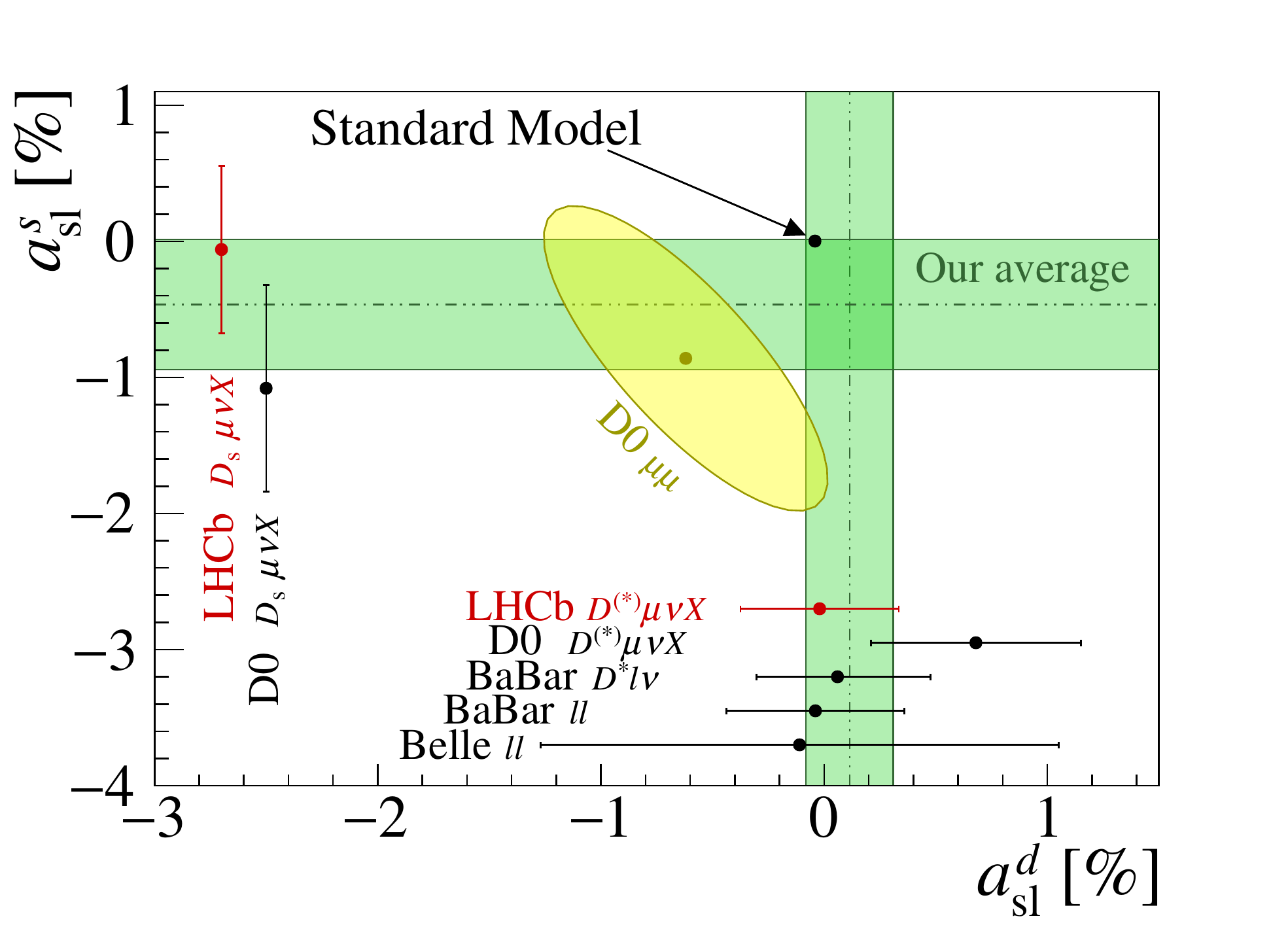}
\caption{ {\it Left:} Fit projections: $B$ decay time and charge asymmetry for the 7~\tev and 8~\tev combined $D^{-}\mu^{+}$ sample. {\it Right:} Overview of the semileptonic asymmetries measurements. The preliminary measurement from the BaBar collaboration: $ a_{\rm sl}^{d} =  (-3.9\pm 3.5({\rm stat}) \pm 1.9({\rm syst})) \cdot 10^{-3}$ \cite{BaBar_asld_new} presented at this conference is included and considered in the average. The bands correspond to the average of the pure \asld and \asls measurements, which excludes the D0 dimuon result.}
\label{fig:fig}
\end{figure}

In this analysis \asld is measured using a sample of $B^{0} \to D^{-}\mu^{+} \nu_{\mu} X$ and  $B^{0} \to D^{*-}\mu^{+} \nu_{\mu} X$ decays. The value of \asld is accessed by measuring the asymmetry between the $D^{(*)-}\mu^{+}$ and $D^{(*)+}\mu^{-}$ final state as function of the \B decay time. The measured final state asymmetry relates to \asld as in Eq. \ref{eq:Ameas}.
The $B^{0}$ production asymmetry is determined together with the \CP violating asymmetry by means of a maximum likelihood fit to the time-dependent decay rates, while the value of the detection asymmetry is measured using independent data-driven techniques.\\
A sample of 1.8 million $B^{0} \to D^{-}\mu^{+} \nu_{\mu} X$ decays and a sample of 0.33 million  $B^{0} \to D^{*-}\mu^{+} \nu_{\mu} X$  decays, with the $D^{*-}$ decaying in $\Dzb \pi$ have been selected to perform the measurement. The Cabibbo favored $D^{-} \to K^{+}\pi^{-}\pi^{-}$ and $ \Dzb \to K^{+}\pi^{-}$ decays are reconstructed, and the direct $CP$ violation in these decays is assumed to be negligible. The different interaction  of a particle with respect to its antiparticle with the detector material and the inefficiency of the detector itself, result in an asymmetry between the \CP conjugated final states. 
For convenience, the ${K^{+}\pi^{-}\pi^{-}\mu^{+}}$ final state is split into two parts, such that the detection asymmetry of the $K\pi$ pair is evaluated separately from the detection asymmetry of the $\mu\pi$ pair.\\
The dominant contribution to the detection asymmetries is given by the $K\pi$ pair, and the largest source is the charged kaon nuclear interaction asymmetry. The detection asymmetry for the $K\pi$ pair is evaluated using prompt $D^{-} \to  K^{+}\pi^{-}\pi^{-}$ and $D^{-} \to  K^{0}\pi^{-}$ decays.  The charge asymmetry of the final state of the first sample contains the detection asymmetry of the  $K\pi$ pair, but it has to be corrected for the $D^{-}$ production asymmetry and the detection asymmetry of the additional pion. The asymmetry of the second control sample, together with the asymmetry due to the neutral kaon interaction ~\cite{LHCb-PAPER-2014-013}, represents  the correction needed. A kinematic re-weighting procedure is needed to match the $K\pi$ pair of the control sample to the $K\pi$ pair kinematics of the signal sample.
The $K\pi$ asymmetry depends on the year and magnet polarity, but is around 1\% on average, with a precision of around 0.1\%.
For the $\mu\pi$ contribution, different methods are used for the different sources of asymmetry. 
Firstly, the events in the signal sample are weighted such that the kinematic distributions of the muon and pion overlap, in order to ensure for the oppositely charged particles no tracking related asymmetries. 
Secondly, the muon trigger and identification asymmetry are evaluated using the same method described for the \asls measurement.
By means of a re-weighting procedure, the kinematics of the control sample is equalized to the signal mode. The
 corrections are at the level of few per mille, depending on the data sample considered. The overall uncertainty is about 0.4\%. Lastly, corrections for the identification efficiency of the pion are taken into account.\\
The first challenge of partial reconstruction is the description of the \B  meson decay time. A $k$-factor is defined as the ratio between the reconstructed and the true momenta in simulated events.
An average \mean{k} factor, depending on the visible mass of the $B$ meson, is used to correct the reconstructed $B$ decay time: $t = L\cdot m_{B} / |\vec{p}| \cdot \mean{k}$, with $L$ the observed flight distance and $p$ the reconstructed momentum of the \B meson, and $m_{B}$ its known mass \cite{PDG2014}.
 The distribution of the $k$-factors, corrected using \mean{k} , is used in the time dependent fit as resolution function, in addition to the usual resolution on the flight distance.
The values of $a_{sl}^d$ and $A_{\rm P}$ are extracted by means of a maximum likelihood fit to the $D^{-}$ or \Dzb mass, the $B$ decay time and the final state particle charge. The $B$ decay time and charge asymmetry fit projections for the combined $D^{-}\mu^{+}$ sample are shown in Fig. \ref{fig:fig} ({\it Left}). The analysis is performed separately for the 7~\tev and 8~\tev center-of-mass energies and for the two magnet polarities of \lhcb \cite{Alves:2008zz}. The results for \asld and $A_{\rm P}$ are then combined.
The largest background is $B^{+} \to D^{(*)-}\mu^{+}X^{+}$ decays, which are around 10\% of the sample. 
 Shapes and fractions are taken from simulation. The production asymmetry for the $B^{+}$ meson, $A_{P}(B^{+}) = (-0.6 \pm 0.6) \%$, is taken from the asymmetry observed in $B^{+} \to J/\psi K^{+}$ decays \cite{LHCb-PAPER-2014-032}, corrected by the $CP$ asymmetry \cite{PDG2014}.
\begin{table}
  \begin{center}
    \vspace{0.1cm}
    \begin{tabular}{l c c c}
      \hline \hline
Source of uncertainty       &  ~~$a_{\rm sl}^d$~~ & $A_{\rm P}({\rm 7~\tev})$ & $A_{\rm P}({\rm 8~\tev})$ \\
 \hline
 Detection asymmetry         &  0.26      &  0.20        &  0.14        \\
 $B^+$ background              &  0.13      &  0.06        &  0.06        \\
 $\Lambda_b$ background              &  0.07      &  0.03        &  0.03        \\
 $B_s^0$ background              &  0.03      &  0.01        &  0.01        \\
 Combinatorial $D$ background &  0.03      &   --         &   --         \\
 $k$-factor distribution     &  0.03      &  0.01        &  0.01        \\
 Decay-time acceptance       &  0.03      &  0.07        &  0.07        \\
 Knowledge of $\Delta m_{d}$           &  0.02      &  0.01        &  0.01        \\
 \hline
 Quadratic sum               & 0.30  & 0.22 & 0.17 \\
 \hline \hline
\end{tabular}
\caption{Systematic uncertainty contributions in [\%], on $a_{sl}^d$ and $A_{\rm P}$ at 7~\tev and 8~\tev center of mass energy. Negligible contributions are marked with "--".}
\label{tab:systematics}
\end{center}
\end{table}
Table \ref{tab:systematics} reports the different sources of systematic uncertainty on \asld and \AP. The leading systematic uncertainty is ascribed to the uncertainty, statistical and systematic, of the measurement of the detection asymmetry.  The $B^{+}$ background currently represents the second largest source of systematic uncertainty for this analysis. The effect of  backgrounds from $\Lambda_{b}^{0}$ and $B^{0}_s$ decays is also evaluated. Systematics related to the decay time description give a smaller contribution. Contributions below the $10^{-4}$ level are considered negligible.
The results obtained for \asld and $A_{\rm P}$ are \cite{LHCb-PAPER-2014-053}:

\begin{equation}
\label{eq:asld_result}
 a_{\rm sl}^{d} =  (-0.02 \pm 0.19({\rm stat}) \pm 0.30({\rm syst}))\%\\
\end{equation}
 \begin{equation}
\label{eq:Ap_result7}
  A_{\rm P}(7~{\tev}) =  (-0.66 \pm 0.26({\rm stat})  \pm 0.22({\rm syst}))\%\\
\end{equation}
  \begin{equation}
\label{eq:Ap_result8}
 A_{\rm P}(8~{\tev}) =  (-0.48 \pm 0.15({\rm stat})  \pm 0.17({\rm syst}))\%
\end{equation}

\section{Conclusions}
The current experimental picture of the semileptonic asymmetries  in neutral $B$ mesons systems is represented in Fig. \ref{fig:fig} ({\it Right}).
Measurements of the $CP$ violation in $B^{0} - \overline{B^{0}}$ and $B_s^{0} - \overline{B_s^{0}}$ mixing performed by the LHCb experiment are reported, focusing on the new result obtained in the $B^{0}$ sector, exploiting the full statistics of Run-I. Both measurements agree with the Standard Model predictions, and they are the current most precise measurements of these quantities from a single experiment.

\Acknowledgements
The author acknowledges the support received from the International Max Planck Research School for Precision Tests of Fundamental Symmetries.

\addcontentsline{toc}{section}{References} 
\bibliographystyle{LHCb} 
\bibliography{eprint}

\end{document}

%% file: my-symbols.tex
\def\myTitle  {Measurement of the semileptonic $C\!P$ asymmetry in $B^0$--$\kern 0.18em\overline{\kern -0.18em B}{}^0$ mixing}

\def\fbar    {{\ensuremath{\kern 0.18em\overline{\kern -0.18em f}{}}}\xspace}
\def\forfbar    {\kern 0.18em\optbar{\kern -0.18em f}{}\xspace}

\def\asld   {\ensuremath{a_{\rm sl}^{d}}\xspace}
\def\asls   {\ensuremath{a_{\rm sl}^{s}}\xspace}
\def\asl    {\ensuremath{a_{\rm sl}}\xspace}
\def\AP     {\ensuremath{A_{\rm P}}\xspace}

\def\AD     {\ensuremath{A_{\rm D}}\xspace}

\newcommand{\gevcNS}{\ensuremath{{\mathrm{Ge\kern -0.1em V\!/}c}}\xspace}

%% file: lhcb-symbols-def.tex
\def\lhcb {\mbox{LHCb}\xspace}

\def\dzero  {\mbox{D0}\xspace}

\def\MagUp {\mbox{\em Mag\kern -0.05em Up}\xspace}

 \mathchardef\PDelta="7101
 \mathchardef\PXi="7104
 \mathchardef\PLambda="7103
 \mathchardef\PSigma="7106
 \mathchardef\POmega="710A
 \mathchardef\PUpsilon="7107
                  
 \def\PB      {\ensuremath{B}\xspace}                 
                  
 \def\PD      {\ensuremath{D}\xspace}

 \def\PK      {\ensuremath{K}\xspace}

 \def\Pi      {\ensuremath{i}\xspace}

 \def\Ps      {\ensuremath{s}\xspace}

\makeatletter
\ifcase \@ptsize \relax
  \newcommand{\miniscule}{\@setfontsize\miniscule{4}{5}}
\or
  \newcommand{\miniscule}{\@setfontsize\miniscule{5}{6}}
\or
  \newcommand{\miniscule}{\@setfontsize\miniscule{5}{6}}
\fi
\makeatother

\DeclareRobustCommand{\optbar}[1]{\shortstack{{\miniscule (\rule[.5ex]{0.8em}{.18mm})}
  \\ [-.7ex] $#1$}}

\def\squark    {{\ensuremath{\Ps}}\xspace}

  \def\Kbar    {{\kern 0.2em\overline{\kern -0.2em \PK}{}}\xspace}

\def\KorKbar    {\kern 0.18em\optbar{\kern -0.18em K}{}\xspace}

  \def\Dbar    {{\kern 0.2em\overline{\kern -0.2em \PD}{}}\xspace}

\def\DorDbar    {\kern 0.18em\optbar{\kern -0.18em D}{}\xspace}

\def\Dzb     {{\ensuremath{\Dbar{}^0}}\xspace}

\def\B       {{\ensuremath{\PB}}\xspace}
\def\Bbar    {{\ensuremath{\kern 0.18em\overline{\kern -0.18em \PB}{}}}\xspace}

\def\BorBbar    {\kern 0.18em\optbar{\kern -0.18em B}{}\xspace}

\def\Bzb     {{\ensuremath{\Bbar{}^0}}\xspace}

\def\Bd      {{\ensuremath{\B^0}}\xspace}
\def\Bs      {{\ensuremath{\B^0_\squark}}\xspace}
\def\Bsb     {{\ensuremath{\Bbar{}^0_\squark}}\xspace}
\def\Bzsb     {{\ensuremath{\Bbar{}^0_{(\squark)}}}\xspace}

  \def\Y#1S{\ensuremath{\PUpsilon{(#1S)}}\xspace}

\def\Lbar        {{\ensuremath{\kern 0.1em\overline{\kern -0.1em\PLambda}}}\xspace}
\def\LorLbar    {\kern 0.18em\optbar{\kern -0.18em \PLambda}{}\xspace}

\def\to                 {\ensuremath{\rightarrow}\xspace}

\def\CP                {{\ensuremath{C\!P}}\xspace}

\newcommand{\dms}{{\ensuremath{\Delta m_{\squark}}}\xspace}

\def\AT#1     {\ensuremath{A_{\mathrm{T}}^{#1}}\xspace}           

\def\C#1      {\ensuremath{\mathcal{C}_{#1}}\xspace}                       
\def\Cp#1     {\ensuremath{\mathcal{C}_{#1}^{'}}\xspace}                    
\def\Ceff#1   {\ensuremath{\mathcal{C}_{#1}^{\mathrm{(eff)}}}\xspace}        
\def\Cpeff#1  {\ensuremath{\mathcal{C}_{#1}^{'\mathrm{(eff)}}}\xspace}       
\def\Ope#1    {\ensuremath{\mathcal{O}_{#1}}\xspace}                       
\def\Opep#1   {\ensuremath{\mathcal{O}_{#1}^{'}}\xspace}                    

\newcommand{\tev}{\ensuremath{\mathrm{\,Te\kern -0.1em V}}\xspace}
\newcommand{\gev}{\ensuremath{\mathrm{\,Ge\kern -0.1em V}}\xspace}
\newcommand{\mev}{\ensuremath{\mathrm{\,Me\kern -0.1em V}}\xspace}
\newcommand{\kev}{\ensuremath{\mathrm{\,ke\kern -0.1em V}}\xspace}
\newcommand{\ev}{\ensuremath{\mathrm{\,e\kern -0.1em V}}\xspace}
\newcommand{\gevc}{\ensuremath{{\mathrm{\,Ge\kern -0.1em V\!/}c}}\xspace}
\newcommand{\mevc}{\ensuremath{{\mathrm{\,Me\kern -0.1em V\!/}c}}\xspace}
\newcommand{\gevcc}{\ensuremath{{\mathrm{\,Ge\kern -0.1em V\!/}c^2}}\xspace}
\newcommand{\gevgevcccc}{\ensuremath{{\mathrm{\,Ge\kern -0.1em V^2\!/}c^4}}\xspace}
\newcommand{\mevcc}{\ensuremath{{\mathrm{\,Me\kern -0.1em V\!/}c^2}}\xspace}

\def\gsim{{~\raise.15em\hbox{$>$}\kern-.85em
          \lower.35em\hbox{$\sim$}~}\xspace}
\def\lsim{{~\raise.15em\hbox{$<$}\kern-.85em
          \lower.35em\hbox{$\sim$}~}\xspace}

\newcommand{\mean}[1]{\ensuremath{\left\langle #1 \right\rangle}} 

\def\tell1  {TELL1\xspace}
\def\ukl1   {UKL1\xspace}



%% file: econfmacros.tex
\def\beq{\begin{equation}}
\def\eeq#1{\label{#1}\end{equation}}
\def\eeqn{\end{equation}}

\def\beqa{\begin{eqnarray}}
\def\eeqa#1{\label{#1}\end{eqnarray}}
\def\eeqan{\end{eqnarray}}

\let\bar=\overbar

\def\Dslash{\not{\hbox{\kern-4pt $D$}}}
\def\dslash{\not{\hbox{\kern-2pt $\del$}}}

\def\msb{{\bar{\ssstyle M \kern -1pt S}}}

